\documentstyle[sprocl,graphicx]{article}
\begin{document}

\title{Effective Field Theory for Neutron-Deuteron Scattering:
Higher Partial Waves}

\author{F.~Gabbiani\footnote{New permanent address: Department of
Physics and Astronomy, Iowa State University, Ames, IA 50011-3160}}

\address{Department of Physics, Duke University\\Box 90305, Durham, NC
27708-0305, USA\\Email: fg@phy.duke.edu}

\maketitle

We present several theoretical results \cite{fabrizio1} for the phase
shifts of higher partial waves for neutron-deuteron elastic
scatterings in the quartet (J = 3/2) and doublet (J = 1/2)
channels. This is the first case in which EFT has been successfully
applied to these channels in a systematic way. Only the pionless case
is investigated here, reserving the inclusion of pion
exchanges for a later investigation.

\begin{figure}[!t]
  \begin{center}
    \vspace{1ex}\noindent \centerline{\includegraphics*[height=0.40\linewidth,
      angle=90]{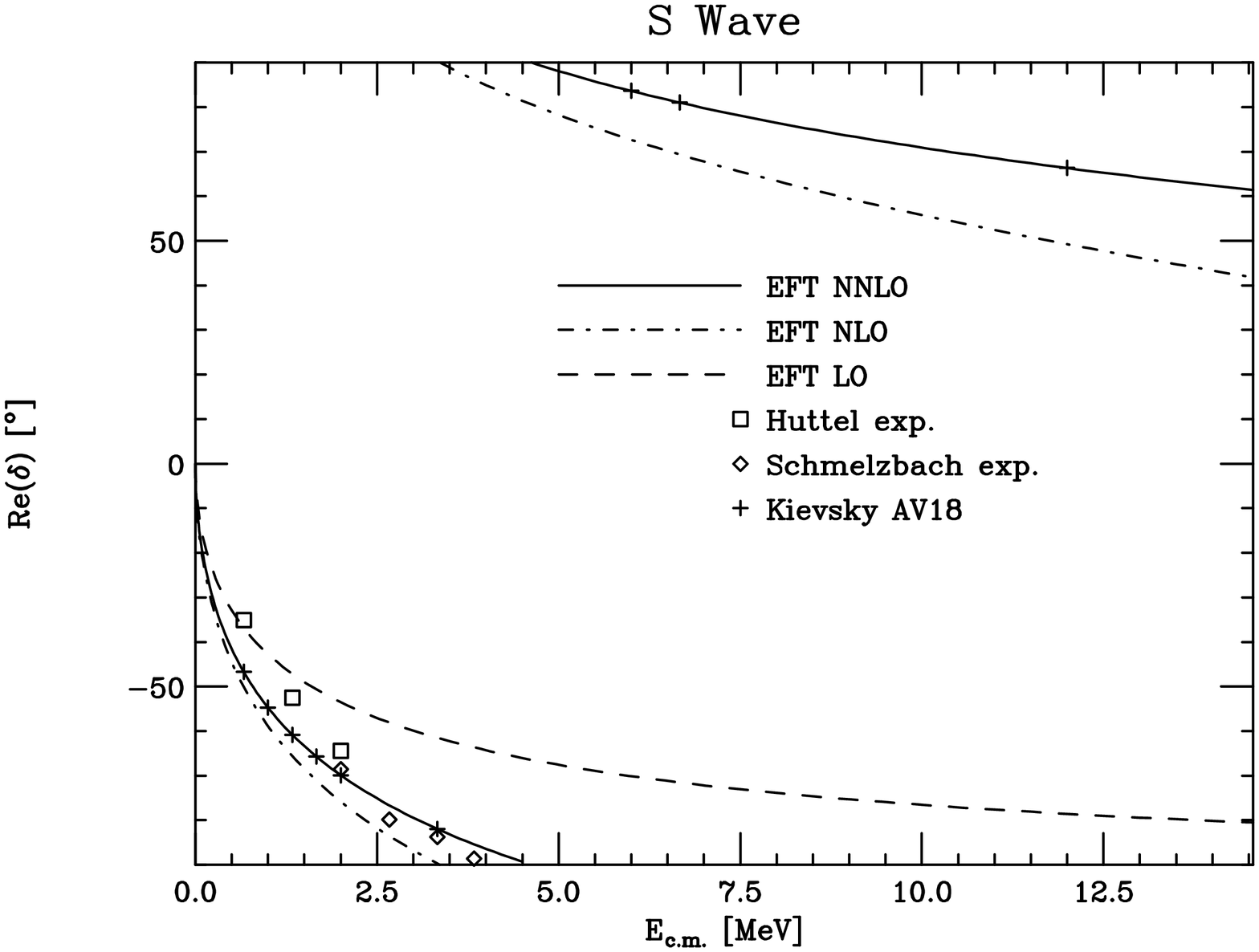} \hfill
      \includegraphics*[height=0.40\linewidth,angle=90]{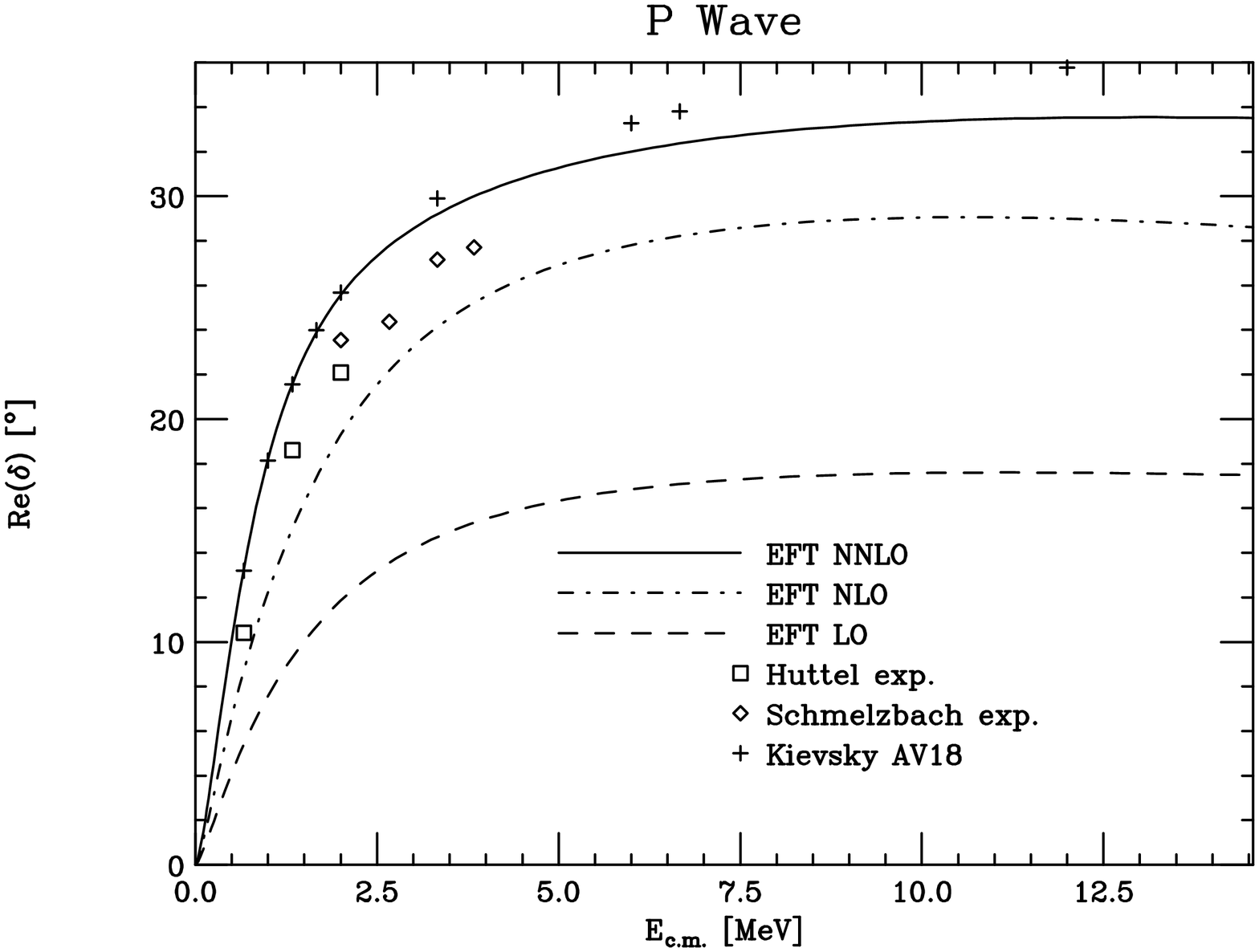} }

    \vspace{1ex}\noindent \centerline{\includegraphics*[height=0.40\linewidth,
      angle=90]{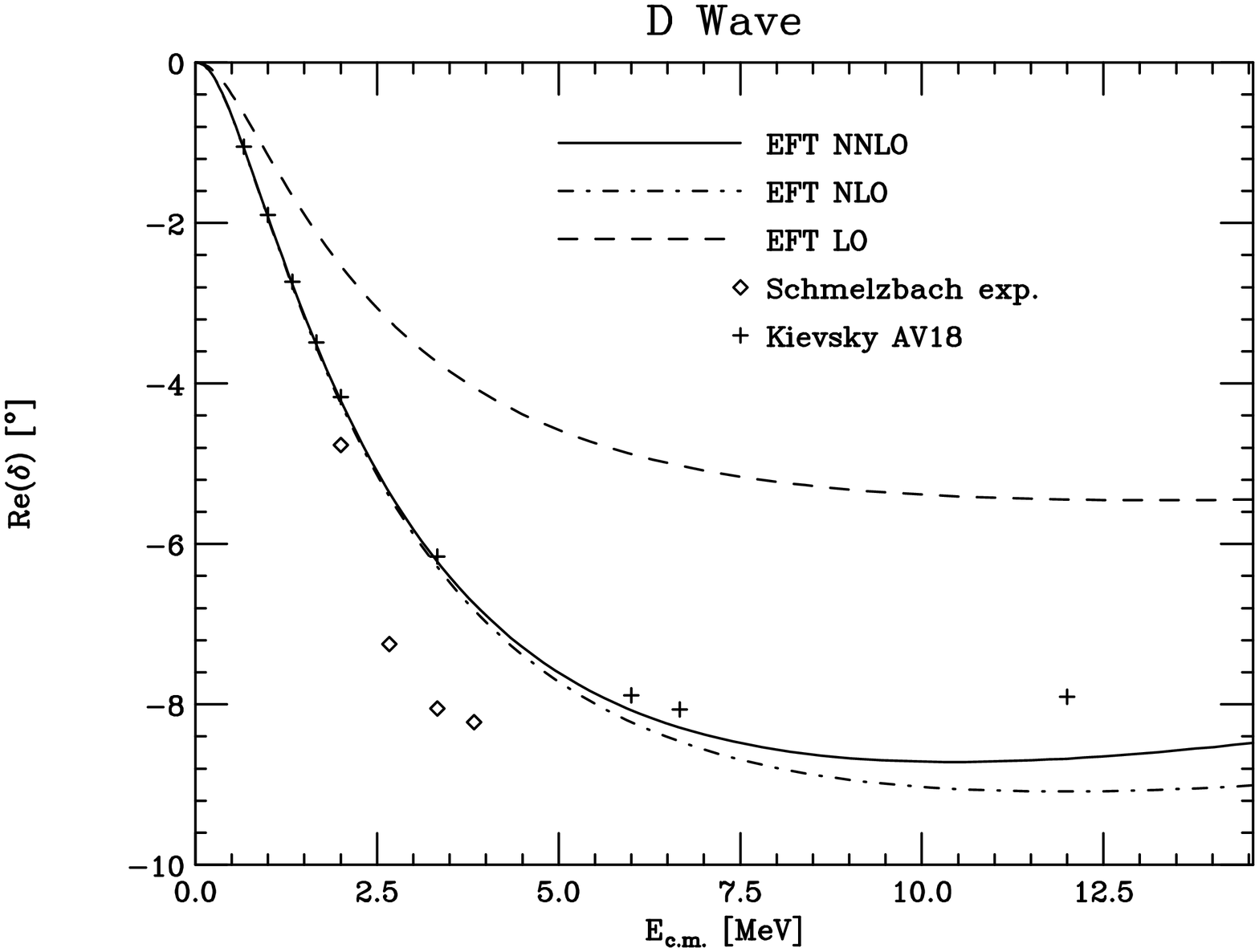} \hfill
      \includegraphics*[height=0.40\linewidth,angle=90]{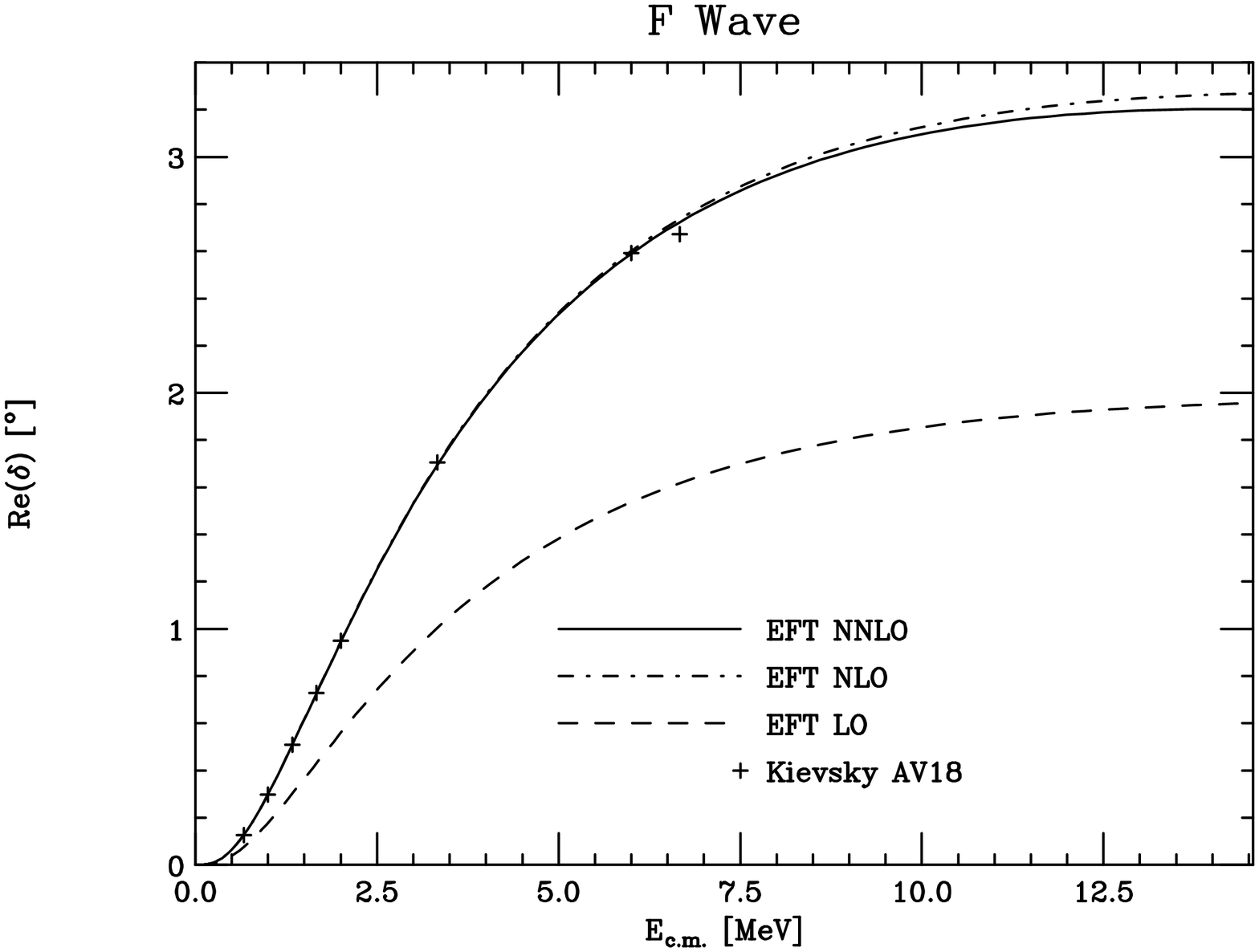} }

     \caption{\sl Real parts of the first four partial waves in
       the quartet channel of \protect$nd$ scattering versus
       the center-of-mass energy.}
    \label{fig:quartet}

    \vspace{1ex}\noindent \centerline{\includegraphics*[height=0.40\linewidth,
      angle=90]{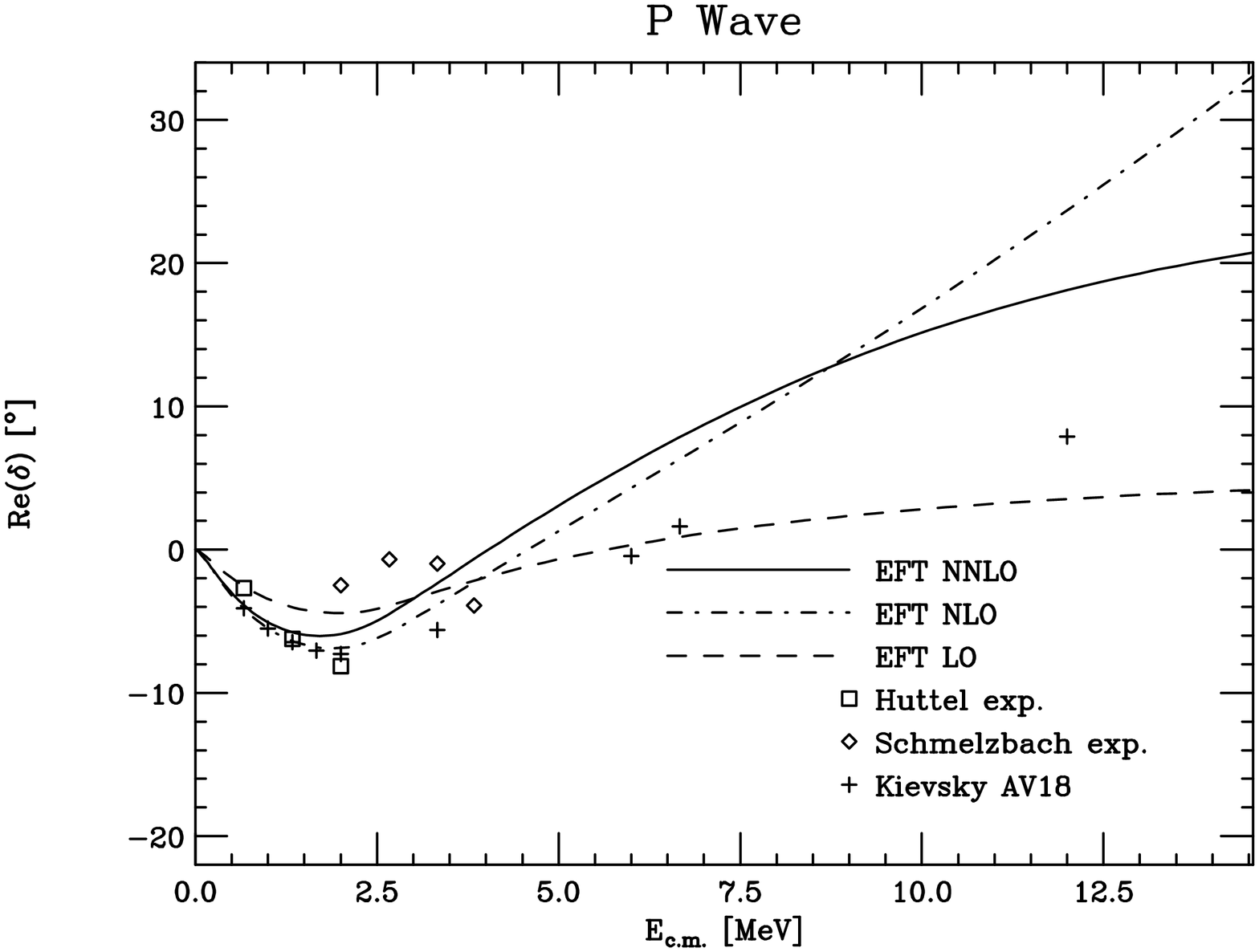} \hfill
      \includegraphics*[height=0.40\linewidth,angle=90]{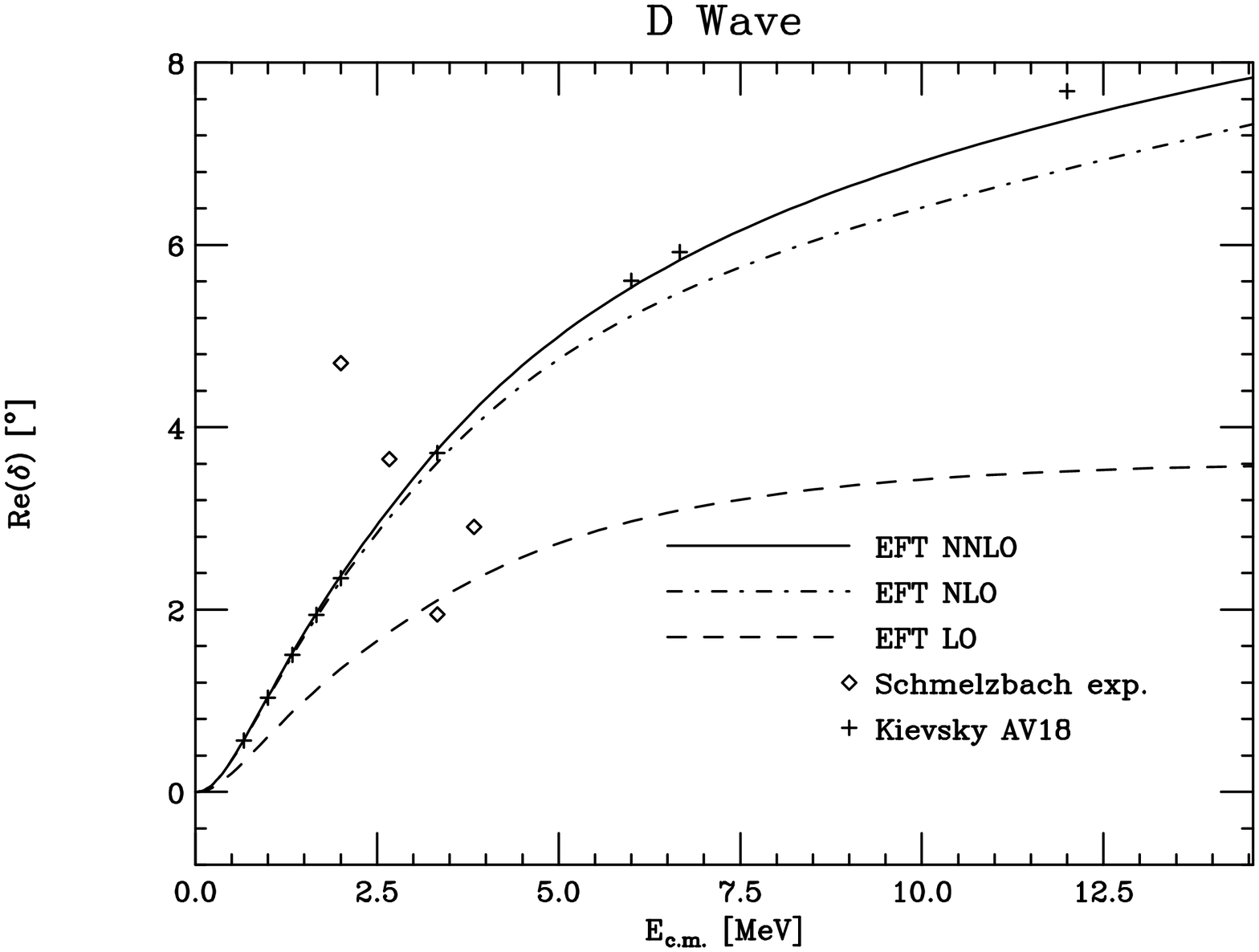} }

    \vspace{1ex}\noindent \centerline{\includegraphics*[height=0.40\linewidth,
      angle=90]{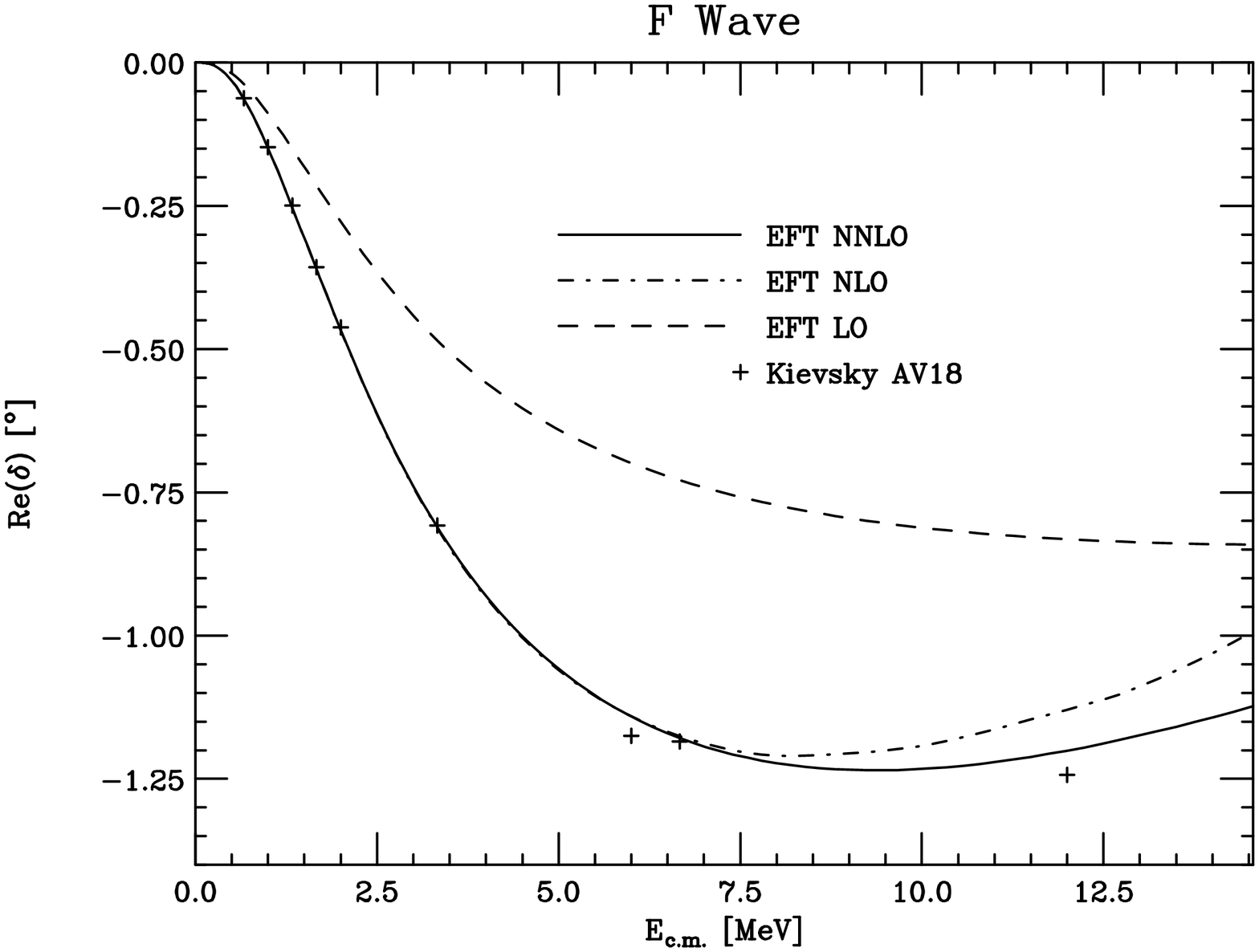} \hfill
      \includegraphics*[height=0.40\linewidth,angle=90]{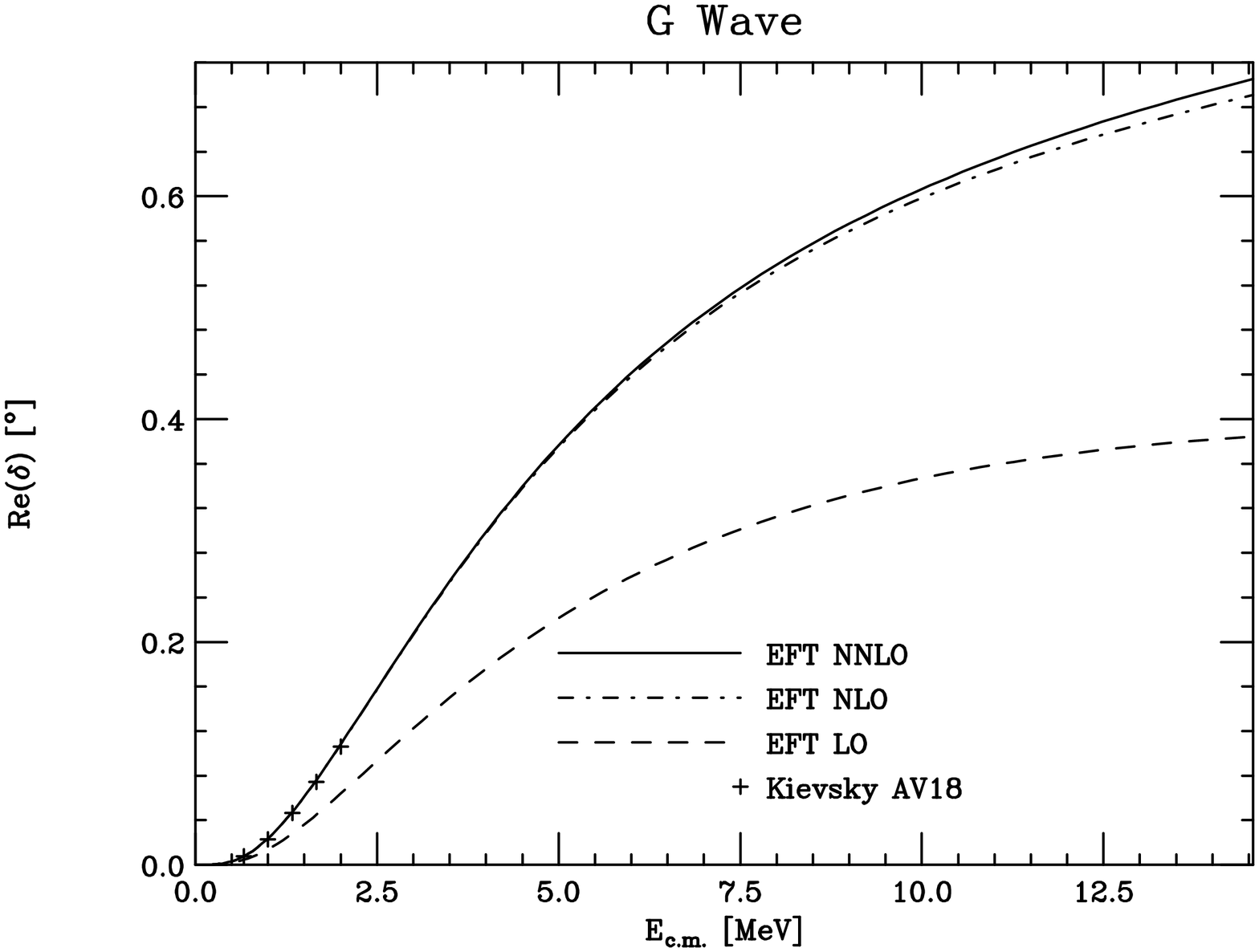} }

    \caption{\sl Real parts of the first four higher partial
      waves in the doublet channel of \protect$nd$ scattering versus the
      center-of-mass energy.}
    \label{fig:doublet}
  \end{center}
\end{figure}

Results for the quartet case are shown in Fig. \ref{fig:quartet}.  The
phase shifts in degrees are plotted vs. the CM energy of the
system. The only experimental results available in the literature are
phase-shift analyses for processes with protons
\cite{fabrizio2}. Yet, their agreement with our theoretical
predictions is excellent, indicating that electro-magnetic effects are
not significant at the energies we study in this work. Potential
models like the ones based on the Bonn and the av$_{14}$ potentials
\cite{fabrizio3} give also similar results, but it is important to
notice that we are able in principle not only to reproduce their
results, but our effective theory framework with its well-defined
power counting will in addition allow us to include higher order
corrections and external currents in a consistent and systematic way.

The doublet case involves the solution of two coupled integral
equations (see for example \cite{fabrizio4}). The S-wave component
relates to the triton bound state, and will be studied in a future
work. We show our results in Fig. \ref{fig:doublet}.

\end{document}